\documentstyle[11pt,paspconf]{article} 

\begin{document} 

\title{Sub-Microarcsecond Astrometry and New Horizons in Relativistic 
Gravitational Physics} 

\author{Sergei M. Kopeikin} 
\affil{Department of Physics and Astronomy, 223 Physics Building, 
University of Missouri-Columbia, Columbia, MO 65211, USA} 

\author{Carl R. Gwinn} 
\affil{Department of Physics, Broida Hall, University of California at 
Santa Barbara, Santa Barbara, CA 93106, USA} 

  \begin{abstract} 
  Attaining the limit of sub-microarcsecond optical resolution will 
  completely revolutionize fundamental astrometry by merging it with 
  relativistic gravitational physics. Beyond the sub-microarcsecond 
  threshold, one will meet in the sky a new population of physical 
  phenomena caused by primordial gravitational waves from early universe 
  and/or different localized astronomical sources, space-time topological 
  defects, moving gravitational lenses, time variability of gravitational 
  fields of the solar system and binary stars, and many others. Adequate 
  physical interpretation of these yet undetectable 
  sub-microarcsecond phenomena can not be achieved on the ground of the 
  "standard" post-Newtonian approach (PNA), which is valid only in the near-zone 
  of astronomical objects having a time-dependent gravitational field. We 
  describe a new, post-Minkowskian relativistic approach for modeling 
  astrometric observations having sub-microarcsecond precision and briefly 
  discuss the light-propagation effects caused by gravitational waves and 
  other phenomena related to time-dependent gravitational fields. The 
  domain of applicability of the PNA in relativistic 
  space astrometry is explicitly outlined. 
  \end{abstract} 

  \keywords{astrometry, gravitational physics, gravitational waves} 

  \section{Theoretical Principles of Relativistic Astrometry} 

  For a long time the basic theoretical principles of general relativistic 
  astrometry in the solar system were based on using the post-Newtonian 
  approximate solution of the Einstein equations (Soffel 1989; 
  Brumberg 1991; 
  Will 1993\noindent
  ). The metric tensor of the
  post-Newtonian 
  solution is an instantaneous function of coordinate time $t$. It depends 
  on the field point, ${\bf x}$, and 
  the coordinates, ${\bf x}_a(t)$, and velocities, ${\bf v}_a(t)$, 
  of 
  the gravitating bodies, and is valid only inside the near zone of the 
  solar system because it involves expansion of retarded field integrals with respect 
  to the small parameter $v_a/c$ (Fock 1959). This expansion restricts the 
  domain of validity for which the propagation of light rays can be 
  considered from the mathematical point of view in a self-consistent 
  manner by the boundary of the near zone. Finding a solution of the 
  equations of light propagation in the near zone of the 
  solar system, for instance, can be achieved by expanding the positions and 
  velocities of the solar system bodies in a Taylor series around some fixed 
  instant of time $t^\ast$, their substitution into the equations of motion of 
  photons, and their subsequent integration with respect to time. Such an 
  approach is theoretically well justified for a proper description of 
  radar (Reasenberg et al. 1979) and lunar laser ranging (Williams et al. 1996) 
  experiments, and interpretation of the Doppler tracking of 
  satellites. However, this static-in-time approach 
  meets with unsurpassable difficulties if one wants to integrate the 
  equations of light propagation from any object lying beyond the limits of 
  the solar system, where the gravitational field can not be considered as 
  static if one decides to search for sub-microarcseconds astrometric 
  effects\footnote{This is because planets are moving around the Sun and, hence, the gravitational field is changing.}. 

  An additional problem arising in expansion about a single instant 
  of time $t^\ast$ is how to 
  determine that fiducial instant of time to which the coordinates and 
  velocities of gravitating bodies should be anchored. The 
  answer is obscured if one works in the framework of the 
  PNA scheme, which disguises the hyperbolic 
  character of the Einstein equations for the gravitational field, and does not 
  admit us to distinguish between advanced and retarded solutions of the 
  field equations (Damour et al. 1991). For this reason, propagation of 
  light rays, which always takes place along the null characteristics 
  of a light cone, differs in the post-Newtonian scheme from that of the 
  gravitational field itself, which propagates (in the framework of this scheme) 
  instantaneously with infinite speed. Thus, the true causal 
  relationship between the position of the light particle and the location of 
  the light-deflecting bodies in the system is violated, which leads to 
  a need for artificial assumptions about the initial values 
  of the positions and velocities of the bodies for integration of equations 
  of light propagation.  One reasonable choice is to fix the 
  coordinates and velocities of the body at the moment of the closest 
  approach of light ray to it. Such an assumption was used 
  by Klioner \& Kopeikin (1992), who 
  proved that it minimizes the magnitude of residual terms of the 
  post-Newtonian solution of the equations of light propagation. 
Among other difficulties in the application of the PNA 
  to solving the problem of propagation of light rays beyond its domain of applicability is the logarithmic divergence of 
  the post-Newtonian metric tensor at large distances from the 
  solar system. Almost all extra-solar luminous objects visible in 
  the sky lie far beyond the allowed distance and, hence, the results of 
  integration of the equations of light propagation from, e.g., stars in 
  our galaxy or extra-galactic objects performed 
  previously by various authors on the premise of the implementation of 
  the post-Newtonian metric tensor can not be considered as rigorous and 
  conclusive, because the magnitude of residual terms of such an integration were never 
  discussed. 

  A real breakthrough in the problem of integration of equations of light 
  propagation in time-dependent gravitational fields has been achieved only 
  recently by Kopeikin et al. (1999), where astrometric and timing effects 
  of the near, intermediate, and far regions of localized gravitational source emitting gravitational waves were precisely 
  calculated and thoroughly discussed on the basis of the post-Minkowskian approximation (PMA) scheme that is free of the drawbacks of the standard PNA. Further progress in solving this problem was made in the paper by 
  Kopeikin \& Sch\"afer (1999), where propagation of light in the field of 
  arbitrary-moving massive monopole particles was described in detail. 
  All possible astrometric effects in VLBI, pulsar timing, Doppler 
  tracking, and so on were considered, including all retardation 
  phenomena in the propagation of a gravitational field from the particles to 
  the photon. We will present a precise description of the 
  gravitomagnetic effects of 
  particles due to their translational and rotational 
  motion in an upcoming publication (Kopeikin \& Mashhoon 2000). The 
  model we are using now is so flexible and physically meaningful that it 
  can be applied for the treatment of any modern astrometric observation as 
  well as for the prediction of numerous physical effects caused by 
  time-dependent gravitational fields including gravitational waves from 
  the early universe, localized sources of gravitational waves like 
  supernova explosions or binary stars, moving and rotating bodies, 
  oscillations of stars, topological defects, and many others. 

  \section{Advanced Relativistic Model of Light Propagation} 

  In this section we briefly describe the relativistic model of light 
  propagation used in our calculations (Kopeikin \& Sch\"afer 1999, 
  Kopeikin et al. 1999, Kopeikin \& Mashhoon 2000) and its application for the calculation of some effects that are important in gravitational 
  physics. 

  The relativistic model of light propagation includes three main constituents: 
  (1) equations of the gravitational field, (2) equations of light 
  propagation, and (3) initial and/or boundary conditions. We formulate 
  the equations of gravitational field in harmonic coordinate system with 
  arbitrary origin in space (for convenience it can be put at 
  the center-of-mass of the source of gravitational field, 
  but this is not essential). 
  We also work in the linear approximation with respect to 
  the universal gravitational constant $G$ without expansion 
  in the velocities of bodies. The metric tensor of the 
  gravitational field can be found immediately from the 
  energy-momentum tensor of the source of gravitational field. 
  The metric tensor is a function of the retarded time argument $s$ related 
  to the running time $t$ and to the coordinates ${\bf x}$ of photon, as well as 
  the retarded 
  coordinates ${\bf x}_a(s)$ of the light-deflecting body by the light-cone equation 
  $s=t-|{\bf x}-{\bf x}_a(s)|$. With the metric tensor in hand we calculate the 
  Christoffel symbols and derive the equations of light propagation, which 
  are second-order ordinary differential equations with retarded 
  argument $s$. Exact solutions of such equations 
  became available only after we found a series of transformations which 
  brought the equations to the integrable form (Kopeikin \& Sch\"afer 1999, 
  Kopeikin et al. 1999). 
  
  Assuming that the light is 
  emitted at a point ${\bf x}_0$ at time $t_0$, and that its direction of propagation 
  is given at past null infinity as a unit vector ${\bf k}$, we integrate the 
  equations of light propagation with no other restrictions. The 
  result is the trajectory of the light ray perturbed by 
  the time-dependent gravitational field, and written as a function of 
  spatial coordinates of the photon, ${\bf x}$, with parametric dependence on 
  the coordinates of the light-deflecting bodies, taken at the retarded 
  time $s$. All predictions of possible relativistic effects (time delay, 
  deflection angle, intensity, polarization angle, etc.) follow directly 
  from the parametric representation of the trajectory (for more details see 
  Kopeikin et al. 1999, Kopeikin \& Sch\"afer 1999, Kopeikin \& Mashhoon 
  2000). 

  \section{Sub-Microarcsecond Astrometry and Relativistic Gravitational Effects} 

  In this section we briefly describe what kind of predictions one can 
  make beyond the sub-microarcsecond level using the aforementioned 
  advanced integration technique. Sub-microarcsecond space-interferometry optical technology along with 
  the advanced post-Minkowskian theory of light propagation in 
  time-dependent gravitational fields, particularly including 
  plane and multipolar gravitational waves, opens outstanding new 
  perspectives for experimental gravitational physics. Here we 
  briefly outline some of the relativistic effects that could be precisely 
  calculated and measured. These effects include: 

  \begin{itemize} 
  \item{1.} Deflection of light in the higher-order PMA
  quadratic in $G$, allowing a static-gravitational-field 
  test of alternative theories of gravity with much better precision 
  than in the weak-gravitational-field regime linear in $G$. 
  \item{2.} Deflection of light caused by gravitational waves emitted by binary 
  stars and other periodic and/or non-periodic sources of gravitational 
  waves, which may allow tests of alternative theories of gravity in the 
  radiative gravitational-wave regime. 
  \item{3.} Various effects caused by a pre-assumed hypothetical difference in the speeds of
propagation of gravity and electromagnetic waves\footnote{This difference may arise, for example, as a result of propagation of light rays through the interstellar medium with the refraction index different from unity or within the framework of an alternative theory of gravity.}. Difference in these 
  speeds would bring about deviation of the gravitational null cone from 
  the electromagnetic one, which can be tested in observations of the 
  deflection of light by giant planets of the solar system or extra-solar gravitational lenses. 
  \item{4.} Specific pattern of proper motions of quasars over the whole sky 
  caused by primordial gravitational waves from the early universe. 
  Certain progress in solving this problem both theoretically and 
  observationally has been achieved by Pyne et al. (1996) and by Gwinn et 
  al. (1997). However, a step over the sub-microarcsecond threshold would 
  make the research in this direction much more profound. 
  \item{5.} Relativistic effect of secular aberration caused by the circular 
  motion of the solar system with respect to the galactic barycenter. 
  Visualization of the secular aberration effect requires calculation of 
  light-ray trajectory in the galactic reference frame with the subsequent 
  relativistic space-time transformation from this frame to the proper 
  reference frame of observer (some details of calculations are given by 
  Kopeikin 1992, Klioner \& Kopeikin 1992). 
  \item{6.} Cosmological gravitational lens effects for deflectors with variable 
  parameters such as mass, spin, quadrupole moment, etc. 
  A super-massive binary black hole in an AGN emitting gravitational waves on 
  a cosmological time-scale is an example of such a source. 
  \item{7.} Testing astrometric effects caused by relic space-time topological 
  defects like vibrating cosmic strings, oscillating cosmic loops, gases of 
  monopoles, textures, etc. 
  \item{8.} Study of limitations on the precision of fundamental astrometric 
  reference frames imposed by gravitational lensing events due to flybys 
  of stars in our galaxy (Sazhin et al. 1998), gravitational waves from an 
  ensemble of galactic binary stars (Kopeikin 1999), and other related 
  phenomena. 
  \end{itemize}
  We are now critically examining these effects in order to work out an appropriate observational strategy which might be used for their detection by VLBI and/or FAME, SIM, and GAIA space-astrometric missions.
  \acknowledgements
  We thank Bahram Mashhoon and Mikhail Sazhin for a critical reading of the manuscript and useful comments.
 
  \end{document}